\documentclass[12pt]{article}

\usepackage{amsmath}
\usepackage{siunitx}
\usepackage{fullpage}
\usepackage{graphicx}
\usepackage[small]{caption}
\usepackage[superscript]{cite}
\usepackage{authblk}

\newcommand{\TitleLine}{Unravelling coherences in the Fenna-Matthews-Olson complex}

\author[,1]{Erling Thyrhaug\footnote{These authors contributed equally to this work.}}
\author[,2]{Roel Tempelaar$^*$}
\author[1]{Marcelo J.~P.~Alcocer}
\author[,1]{Karel \v{Z}\'{i}dek\footnote{Current address: TOPTEC, Institute of Plasma Physics, Academy of Sciences of the Czech Republic, Za Slovankou 1782/3, 182 00 Prague 8, Czech Republic}}
\author[3]{David B\'{i}na}
\author[4]{Jasper Knoester}
\author[4]{Thomas L.~C.~Jansen}
\author[,1]{Donatas Zigmantas\footnote{donatas.zigmantas@chemphys.lu.se}}

\affil[1]{\small Department of Chemical Physics, Lund University, P.O. Box 124, 22100 Lund, Sweden}
\affil[2]{\small Department of Chemistry, Columbia University, 3000 Broadway, New York, New York 10027, USA}
\affil[3]{\small Biology Centre CAS, Brani\v{s}ovsk\'{a} 31, and Faculty of Science, University of South Bohemia, Brani\v{s}ovsk\'{a} 1760, 370 05 \v{C}esk\'{e} Bud\v{e}jovice, Czech Republic}
\affil[4]{\small University of Groningen, Zernike Institute for Advanced Materials, Nijenborgh 4, 9747AG Groningen, The Netherlands}

\date{}

\begin{document}

\title{\TitleLine}

	\maketitle
	\begin{abstract}
		
	The idea that excitonic state (electronic) coherences are of fundamental importance to natural photosynthesis gained popularity when, a decade ago, slowly dephasing quantum beats were observed in the two-dimensional electronic spectra of the Fenna-Matthews-Olson complex at 77\,K. These were assigned to quantum superpositions of excitonic states; a controversial interpretation, as the spectral linewidths suggested fast dephasing arising from strong interactions with the environment. While it has been pointed out that vibrational motion produces similar spectral signatures, concrete assignment of these coherences to distinct physical processes is still lacking. 
	Here we revisit the coherence dynamics of the Fenna-Matthews-Olson complex using polarization-controlled two-dimensional electronic spectroscopy, supported by theoretical modelling. We show that the long-lived quantum beats originate exclusively from vibrational coherences, whereas electronic coherences dephase entirely within \SI{240}{\femto\second} even at 77\,K -- a timescale too short to play a significant role in light harvesting. Additionally, we demonstrate that specific vibrational coherences are excited via vibronically coupled states. The detection of vibronic coupling indicates the relevance of this phenomenon for photosynthetic energy transfer.

	\end{abstract}

\section*{} 

	Through billions of years of evolution, nature has found a solution for the efficient harvesting of sunlight in the form of densely packed pigments embedded in protein environments\cite{Cogdell_06a, Scholes_11a}. In aiming to understand the functionality of these complexes, particular attention has been paid to the Fenna-Matthews-Olson (FMO) complex\cite{Fenna_75a} -- a small protein homo-trimer situated between the chlorosome antennae and photosynthetic reaction centre (RC) of green sulphur bacteria \cite{Hauska2001_GSB_RC, dostal_situ_2016}. The historic interest in FMO has been due to its water solubility, the early resolution of its crystal structure \cite{Tronrud2009, Ben-Shem}, and its relative structural simplicity (Fig.~\ref{fig:fmo}a). Together, these properties make the complex experimentally accessible whilst still being simple enough to allow for detailed theoretical work. The assumption has thus been that it could serve as an exemplary system for unravelling the mechanisms underlying photosynthetic light harvesting.

	Following its structural determination, decades of experimental and theoretical studies have resulted in detailed descriptions of the excitonic structure and the energy transfer dynamics \cite{Vulto_98a, Brixner_05a, Cho_05a, Adolphs_06a, Savikhin_1994, Savikhin_1998} in FMO. More recently, it has been studied by two-dimensional electronic spectroscopy (2DES) \cite{Hybl_98a, Jonas_03a} , which has enabled direct tracking of the energy flow in both the free \cite{Brixner_05a, Thyrhaug_16a} and \textit{in situ} \cite{dostal_situ_2016} complexes. 

	The prevailing paradigm for energy transfer in weakly or intermediately coupled systems such as FMO has been one based on incoherent exciton `hopping'. Models based on this picture have been used with great success to quantitatively explain energy transfer in a wide variety of photosynthetic complexes, including FMO \cite{Vulto_99}. In 2007, a strongly contrasting picture received significant attention when Fleming, Engel, and co-workers reported long-lived quantum beats (QBs) in the 2DES signals of FMO at \SI{77}{\kelvin} \cite{Engel_07a, Panitchayangkoon_10a} and attributed them to coherent superpositions of excitonic states. Such excitonic coherences had already been identified in 1997 \cite{Savikhin_97a}, but were however observed to dephase with a time constant of less than \SI{180}{\femto\second} even at \SI{19}{\kelvin}. Subsequent observations of similar QBs in the 2DES signals from other photosynthetic complexes suggested that such coherence dynamics could be crucial to photosynthetic function \cite{ Collini_10a, Schlau-Cohen_12a, Harel_12a}. 

	Since its proposition, this coherent excitonic interpretation has been highly controversial, since the broad spectral lines of light-harvesting complexes imply fast dephasing due to strong coupling of electronic states to the environment. For FMO in particular, estimates of the optical dephasing times based on absorption linewidths barely exceed \SI{100}{\femto\second} even at \SI{77}{\kelvin} \cite{Wendling_00a} (Fig.~\ref{fig:fmo}b). To overcome this apparent contradiction, correlated site-energy fluctuations for the protein-bound pigments were proposed \cite{Engel_07a, Lee_07a}. Subsequent simulations failed however to identify any such `protection' of coherences \cite{Olbrich_2011, Shim_2012}. 

	More recent studies have indicated that the observed spectroscopic signals could in fact be explained by vibronic coupling of excited states. Several excitonic energy gaps in FMO are found around 150 to 240\,cm$^{-1}$ -- a range where a number of vibrational modes are also present \cite{Ratsep_07a}. By explicitly incorporating such modes into a vibronic exciton model, it was shown that long-lived coherences of mixed vibronic character could be produced in the excited state \cite{Christensson_12a}. Later it was demonstrated that ground-state vibrations can also contribute to signals similar to those observed in FMO \cite{Tiwari_13a} when excited via vibronically coupled transitions. A subsequent study incorporating the entire FMO subunit concluded that ground-state coherences in fact can dominate the 2DES signal \cite{Tempelaar_14a}, however no direct experimental evidence to support this has emerged so far.

	In this study we apply phase-sensitive analysis of the QBs in data obtained from two distinct sequences of polarized pulses to characterize coherences in the FMO complex at \SI{77}{\kelvin}. We clearly distinguish short-lived excitonic coherences and long-lived vibrational coherences both in ground and excited states.

\section*{Results} 

\subsubsection*{Structure and Absorption of the FMO complex}
\begin{figure*}[!t]
	\includegraphics[width=8.8cm]{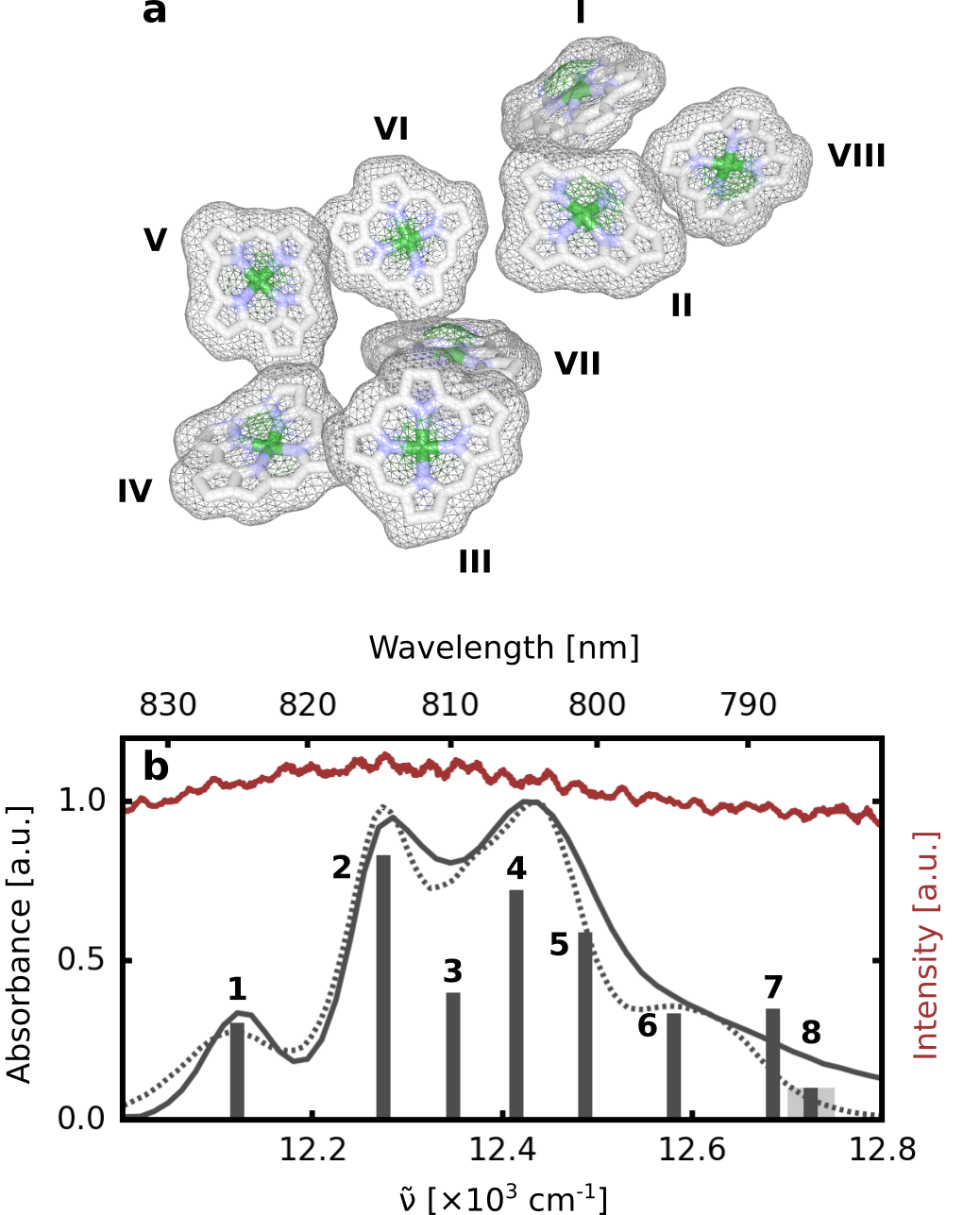}
	\centering
	\caption{\textbf{Structure and absorption of the FMO complex (a)} Structural arrangement of the bacteriochlorophyll a pigments in the FMO complex (data from Ref.~\cite{Tronrud2009}) with site numbering according to Fenna. \textbf{(b)} Experimental (solid) and calculated (broken) absorption spectra of FMO at \SI{77}{\kelvin}. Experimentally determined \cite{Thyrhaug_16a} exciton energies (sticks) and laser spectrum (red) used in 2DES experiments are also shown.}
	\label{fig:fmo}
\end{figure*}

The initial crystallographic work on the FMO complex found the protein subunits to contain well-defined structures of seven bacteriochlorophyll~\textit{a} (BChl) pigments \cite{Fenna_75a}. This was however later amended to accommodate the discovery of a loosely-bound eighth BChl in each subunit \cite{Tronrud2009} (Fig.~\ref{fig:fmo}a). Spectroscopic signatures of the eighth BChl were indeed found in our preceding FMO study \cite{Thyrhaug_16a}, and it was thus presumed that the isolated FMO complexes investigated here also contained eight BChl molecules. The previously extracted exciton energies and \SI{77}{\kelvin} absorption spectrum of the FMO complex isolated from green sulfur bacteria \textit{Chlorobium tepidum} are shown in (Fig.~\ref{fig:fmo}b).

\subsubsection*{Coherence Signals in Polarization-Controlled Two-Dimensional Electronic Spectroscopy}

	The 2DES technique \cite{Jonas_03a} and our fully non-collinear implementation of it \cite{Augulis_11a} have been detailed previously. The recorded dataset appears as a sequence of two-dimensional maps where the complex emitted field, $E^{(3)}(\tilde{\nu}_1,t_2,\tilde{\nu}_3)$, is displayed as a function of excitation and detection energies (proportional to wavenumbers $\tilde{\nu}_1$ and $\tilde{\nu}_3$ respectively) and evolves with the population time $t_2$. As QBs appear along $t_2$, a Fourier transform over the population time allows convenient identification of coherences on excitation/detection coordinates. 
	We refer to the resulting maps as $\tilde{\nu}_{2}$ oscillation maps. 

	2DES datasets are information dense as they contain the entire third-order response of the system. When applied to multi-chromophore systems, this density often leads to problematic spectral congestion. Since the measured signals are dependent on both the relative angles between transition dipole moments and the relative polarization angles of the incident laser pulses, polarisation spectroscopy techniques can be used to alleviate such congestion. These are particularly powerful approaches in fully non-collinear 2DES geometries, where one can control the polarizations of all incident pulses and the detected signal.

	Most reported studies on FMO\cite{Engel_07a, Panitchayangkoon_10a} and other photosynthetic complexes\cite{Collini_10a, Harel_12a, Fuller_14a, Romero_14a} have exclusively employed a series of parallel-polarized pulses -- denoted here as `all-parallel' (AP) or $\langle0^\circ,0^\circ,0^\circ,0^\circ\rangle$. While this sequence typically yields the strongest signal, it preferentially generates signals originating from interactions between parallel dipoles. As a consequence, coherence dynamics in these experiments are dominated by \textit{intramolecular} vibrational motion.
	The transitions generating these coherences do not couple states on different pigments and are not expected to impact energy mobility. 

\begin{figure*}[!t]
	\includegraphics[width=17cm]{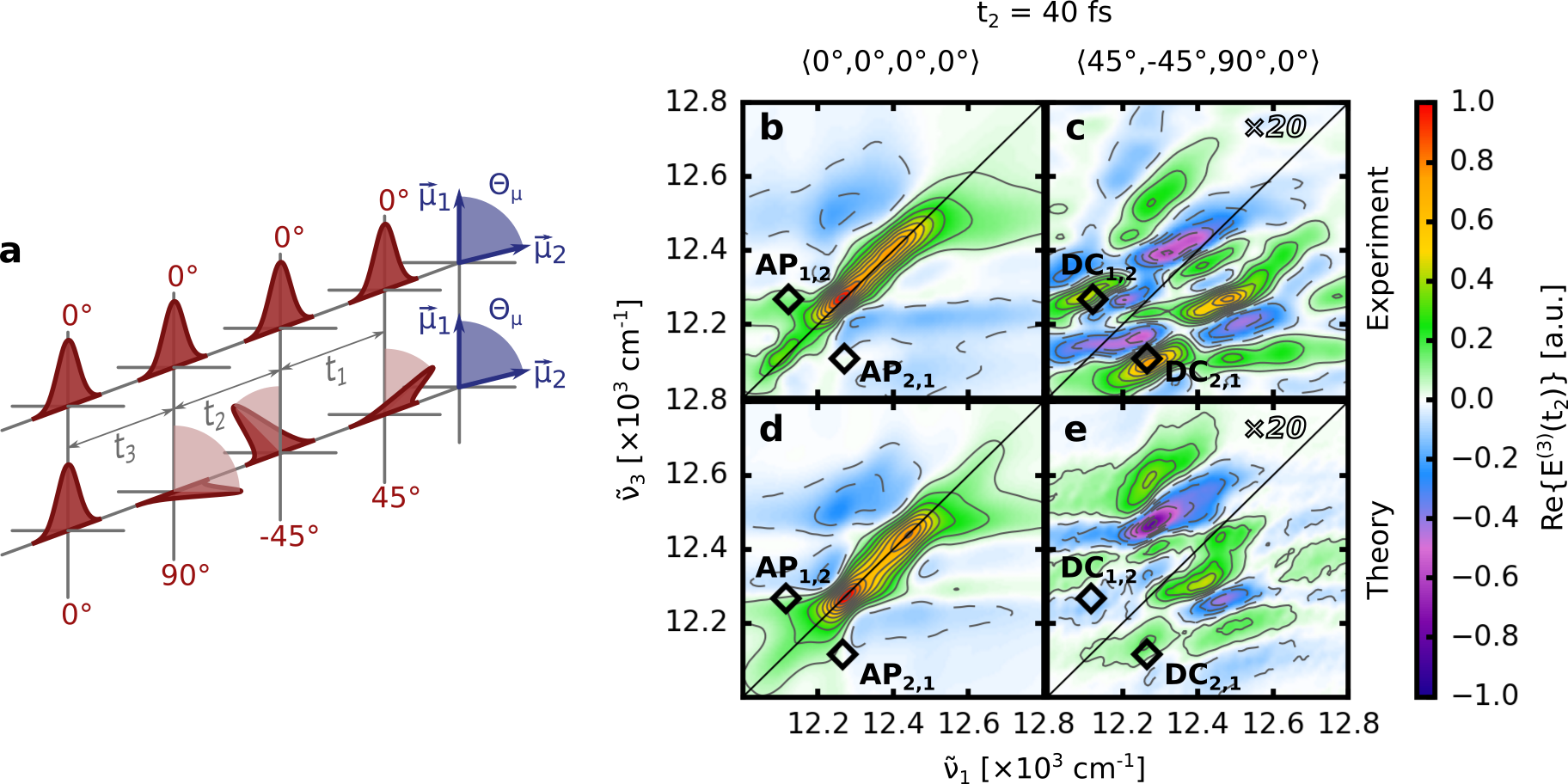}
	\centering
	\caption{\textbf{Polarization-controlled 2DES of the Fenna-Matthews-Olson complex. (a)} Schematic representation of the all-parallel pulse sequence (top) which favours interaction pathways involving parallel transition dipoles ($\vec{\mu}_1$ only), and the double-crossed pulse sequence (bottom) which favours pathways involving non-parallel transition dipoles ($\vec{\mu}_1$ and $ \vec{\mu}_2$). (Right) real part of the rephasing 2D spectra at $t_2 = 40$\si{\femto\second}. Experimental (\textbf{b, c}) and theoretical (\textbf{d, e}) spectra resulting from AP (\textbf{b, d}) and DC (\textbf{c, e}) pulse sequences are shown. The spectra are normalized to the exciton 2 diagonal peak of the AP spectra and the DC spectra are scaled by a factor of 20 for clarity. Markers \textbf{AP} and \textbf{DC} denote the exciton 1--2 and 2--1 cross-peak positions in the appropriate spectra at which the dynamics presented in Fig.~\ref{fig:dyn} are taken.}
	\label{fig:spec}
\end{figure*}

	To suppress such \textit{intramolecular} signals and allow the signatures of coherence across multiple pigments to emerge, we apply a sequence of two perpendicularly-polarized pulse pairs -- denoted here as `double-crossed' (DC) or $\langle45^\circ,-45^\circ,90^\circ,0^\circ\rangle$ (Fig.~\ref{fig:spec}a). First applied in 2D vibrational spectroscopy\cite{Hochstrasser_01a, Zanni_01a} and later in 2DES\cite{Schlau-Cohen_12a, Westenhoff_12a}, it suppresses both all non-coherence signals (\textit{e.g.} population dynamics), and coherence signals involving interactions with pairwise parallel transition dipoles. Thus, signals from localized vibrational modes are suppressed, while signals from \textit{e.g.} \textit{intermolecular} electronic coherences remain. 

	In coupled multi-chromophore systems, certain linear combinations of vibrational modes -- as viewed in a delocalized vibrational basis -- can also contribute to the DC signal. More specifically, this occurs when a coherence is generated through transitions to (vibronically) mixed excited states (see Supplementary Fig.~6). As investigated in detail by Tiwari \textit{et al.} \cite{Tiwari_13a}, this results in the electronic character of the excited states taking on a vibrational coordinate dependence, effectively causing the transition polarization to oscillate with the vibrational frequency. Polarization spectroscopy is sensitive to such an effect and so provides a uniquely powerful tool to analyse the nature of coherences. Of particular relevance to FMO, vibronically coupled states are thus revealed in polarization spectroscopy through the presence of vibrational contributions in the DC signal.

	To provide support to the experimental assignment of the coherence signals, we simulate the time evolution of the FMO subunit using a vibronic exciton model, calculating a polarization-resolved 2D spectrum at each population time step. The model (see Ref.~\cite{Tempelaar_14a} and Supplementary Section~1 for details) allows explicit inclusion of a Raman-active vibrational mode for each bacteriochlorophyll (BChl), which was parametrized with a Huang-Rhys factor of 0.02 and a wavenumber of \SI{160}{\per\centi\metre}. The weakly coupled eighth BChl and the weak interactions between BChls on different trimeric units were neglected, and the calculations were thus based on the 7-site electronic Hamiltonian obtained in earlier studies \cite{Vulto_98a}. The calculated absorption spectra of FMO are shown in Fig.~\ref{fig:fmo}b.

\subsubsection*{Two-Dimensional Spectra of FMO}

	In order to characterize the coherence dynamics in FMO with high spectral resolution, polarization-controlled 2DES experiments were performed at \SI{77}{\kelvin}, scanning the population time $t_2$ to \SI{2}{\pico\second}. The dramatic difference in spectral structure between the pulse polarization sequences is clear from inspection of the spectra in Fig.~\ref{fig:spec}b and c, where the real (absorptive) part of representative ($t_2 = 40$\si{\femto\second}) rephasing 2D spectra are shown (total 2D spectra are shown in Supplementary Fig.~1).

	The AP spectra (Fig.~\ref{fig:spec}b) are dominated by peaks on the diagonal, which can be readily associated with features in the absorption spectrum (Fig.~\ref{fig:fmo}b). The patterns of off-diagonal features, on the other hand, reveal correlations between these bands. The excitonic structure and relaxation pathways emerging from analysis of the time-evolution of 2D spectra have been detailed elsewhere\cite{Thyrhaug_16a}. 

	In the DC spectra (Fig.~\ref{fig:spec}c), signals from population-dynamics are suppressed, and the remaining signals are `running waves' across the 2D map of alternating negative and positive features. Correspondingly, the time-evolution of AP and DC spectra also differ radically; while multi-exponential dynamics dominate the AP spectra with only weak QBs observable ($< 5\%$ total amplitude), the DC spectra are essentially purely oscillatory (Supplementary Fig.~2).

Despite the great complexity in modelling the FMO spectral response, the simulated spectra (Fig.~\ref{fig:spec}d--e) show good agreement with experimental data for both pulse polarization sequences. This indicates that the chosen parametrization of the model captures essential features of the complex.

\subsubsection*{FMO Quantum Beats}

\begin{figure}[!t]
	\centering
\includegraphics[width=8.8cm]{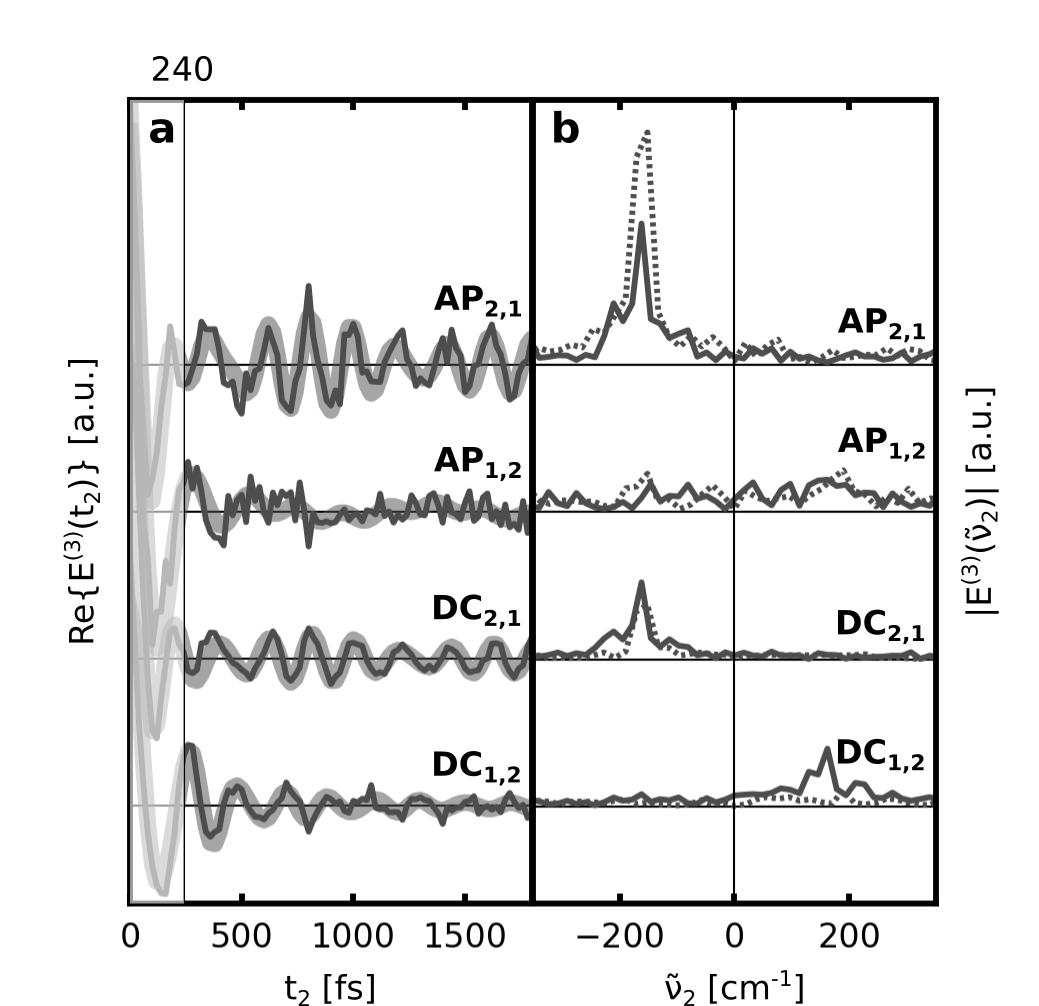}
	\caption{\textbf{Selected quantum beats in FMO. (a)} Real part of experimental rephasing $t_2$ dynamics (black) at the cross-peak locations labelled in Fig.~\ref{fig:spec} after subtraction of multi-exponential dynamics. Individual time-domain fits (grey) corresponding to the oscillatory terms in Eq.~\ref{eq:fit} are overlaid onto each trace (fit parameters can be found in Supplementary Table~1). The traces are vertically offset for clarity. The traces shown here are \textit{not} scaled. \textbf{(b)} Fourier transform amplitudes of the experimental data shown in \textbf{(a)} and theoretical vibronic exciton model data extracted from the same points (broken).}
\label{fig:dyn}
\end{figure}

	In both AP and DC experiments, the areas in the vicinity of the cross-peaks connecting the two lowest-energy excitons (located at $\bar{\nu} = 12,120$ and $12,270$\,cm$^{-1}$) show particularly prominent QBs. These areas are labelled AP$_{1,2}$, AP$_{2,1}$, and DC$_{1,2}$, DC$_{2,1}$ in Fig.~\ref{fig:spec} for the AP and DC pulse sequences respectively. In order to quantitatively study the QBs, the (complex) $t_2$ dynamics at these spectral locations were fitted with a sum of damped complex oscillations and exponential decays
\begin{align}
	E^{(3)}(t_2) = \sum_{m} A^{QB}_m e^{-(t_2/\tau^{QB}_m) - i\omega_m t + \phi_m} \nonumber \\
	+ \sum_{n} (A^{Re}_n e^{-t_2/\tau^{Re}_n} + iA^{Im}_n e^{-t_2/\tau^{Im}_n}) 
	\label{eq:fit}
\end{align}
where $A$ are amplitudes, $\omega$ -- (positive or negative) angular frequencies, $\phi$ -- phases, $\tau$ -- decay time constants, and $m,n = 1,2,3$. To facilitate analysis of the QBs, the experimental $t_2$ dynamics are presented in Fig.~\ref{fig:dyn}a after subtraction of the multi-exponential fit components associated with population dynamics (second terms in Eq. \ref{eq:fit}). The corresponding oscillatory components of the fits (first terms in Eq. \ref{eq:fit}) are also overlaid (full fits showing the imaginary [dispersive] part of the signal and all fit parameters are presented in Supplementary Section~4).  

The initial cross-peak dynamics are dominated by a high-amplitude QB component which decays completely within 240\,fs. While an accurate frequency cannot be extracted due to the sub-cycle dephasing times, the approximate oscillatory periods are consistent with the energy splitting between the two lowest-energy excitons. 
The dephasing time of 50--150\,fs (see Supplementary Table~1), is further consistent with the decay of excitonic coherence observed in earlier cryogenic pump-probe experiments \cite{Savikhin_97a}. We thus attribute the rapidly dephasing QBs to coherent superpositions of excitonic states. It is noteworthy that no substantial low frequency coherences were observed in a very recent 2DES study of FMO at ambient temperatures. This was interpreted as excitonic dephasing occurring on a timescale of 60\,fs\cite{Duan_2017}.

The current investigation is however primarily concerned with the interpretation of slowly dephasing coherences ($\tau^{QB} > 1$\si{\pico\second}). As such, we henceforth only consider the late-time dynamics ($t_2 > 240$\si{\femto\second}), where a number of lower-amplitude QBs persist for several picoseconds. To elucidate the origin of these QBs, it is instructive to consider them in the frequency (or wavenumber $\tilde{\nu}_2$) domain. The complex Fourier transform amplitudes of the experimental and theoretical late-time dynamics are presented in Fig.~\ref{fig:dyn}b,  where well-defined peaks appear at positive or negative wavenumbers, depending on the spectral location. This discrimination into positive and negative wavenumbers originates from the separation of signals evolving with opposite phase in the complex plane -- i.e. as either $e^{-i \omega t_2}$ or $e^{+i \omega t_2}$. This separation proves highly useful when assigning coherence signals to specific physical processes \cite{Li_2013,Seibt_2013,Butkus_2017,Thyrhaug_2017}.

In the below-diagonal cross-peak area investigated in earlier work\cite{Engel_07a, Panitchayangkoon_10a} and labelled here as AP$_{2,1}$ and DC$_{2,1}$, the dynamics are dominated by \SI{-170}{\per\centi\metre} and \SI{-210}{\per\centi\metre} modes exhibiting dephasing times of $\sim$\SI{2}{\pico\second}. Similar wavenumber vibrational modes also appear in the hole-burning and fluorescence line-narrowing spectra of FMO \cite{Ratsep_07a}, and both corresponding energies are close to several excitonic gaps (see Fig.~\ref{fig:fmo}b). We do not however observe any higher wavenumber features with substantial amplitude, in direct contrast to Ref. \cite{Engel_07a}, where coherences were reported at $\sim$350 and $\sim$\SI{500}{\per\centi\metre}.

The behaviour of QBs at AP$_{2,1}$ and DC$_{2,1}$  and their relative amplitudes are well reproduced by the theoretical model, whose frequency domain response is also presented in Fig.~\ref{fig:dyn}b.  In particular, the simulations clearly capture the intensity suppression imposed by the DC sequence. Not that for each BChl only the \SI{160}{\per\centi\metre} mode was included in the model.

The above-diagonal cross-peak, DC$_{1,2}$, on the other hand is dominated by positive wavenumber QBs. While the wavenumbers extracted from time domain fits are identical to the wavenumbers seen below diagonal, the dephasing times of these QBs are noticeably shorter, with the average dephasing constant of  \SI{170}~and \SI{210}{\per\centi\metre} modes corresponding to $\sim$\SI{570}{\femto\second} (see Fig.~\ref{fig:dyn} and Supplementary Table~1). These QBs are not accurately reproduced in the simulation, suggesting they arise from a physical process not explicitly captured by the vibronic model in its current form. We discuss the origin of these QBs in the following sections.

\subsubsection*{Oscillation Maps}

\begin{figure*}[!t]
	\includegraphics[width=17cm]{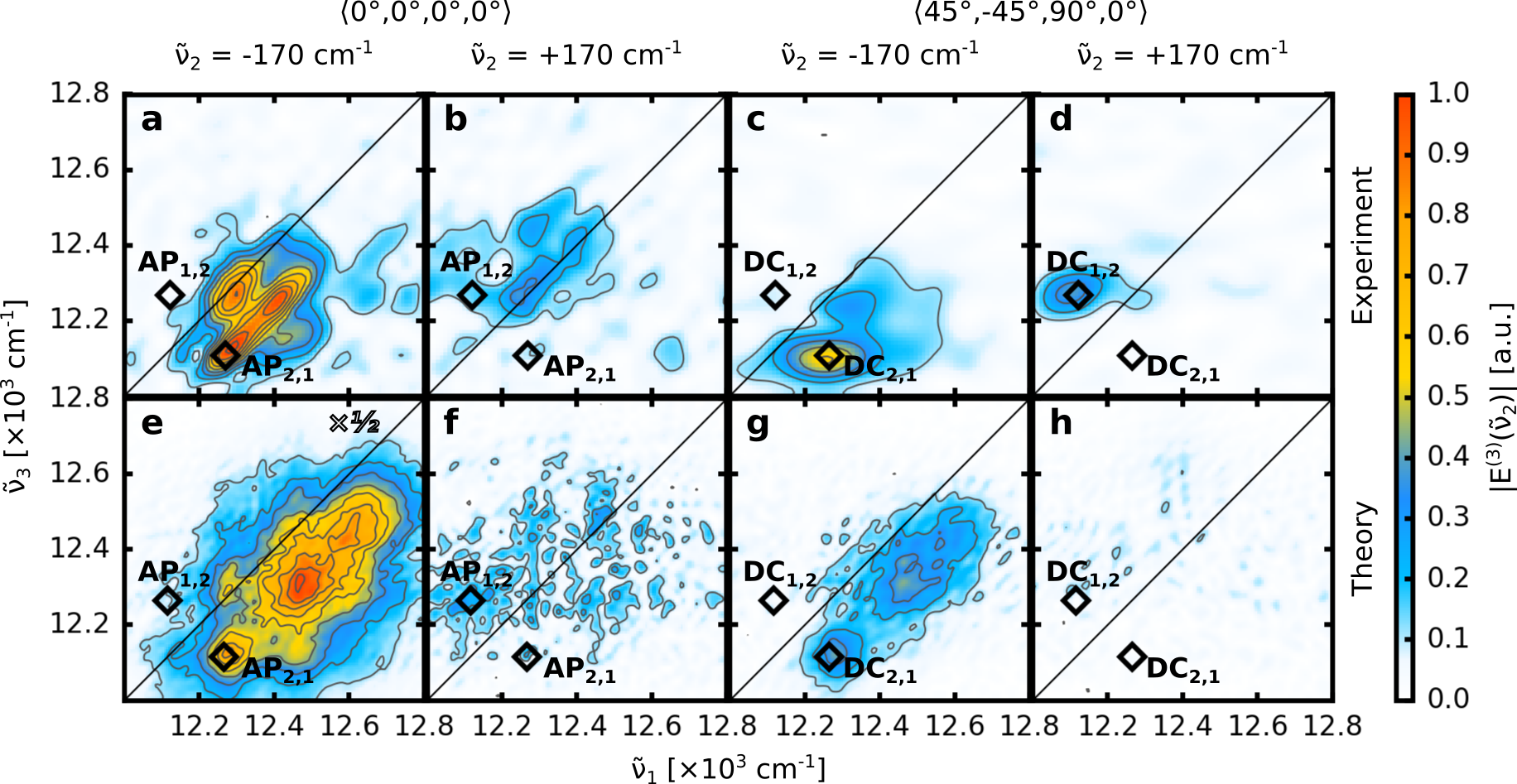}
	\centering
	\caption{\textbf{Oscillation maps.} Fourier amplitude maps at $\pm$ \SI{170}{\per\centi\metre} obtained by Fourier transformation of the 2D datasets along $t_2$ after subtraction of multi-exponential population dynamics. Maps for the experimental (top row, \textbf{a--d}) and theoretical (bottom row, \textbf{e--h}) datasets under the all parallel (left, \textbf{a,b,e,f}) and double-crossed (right, \textbf{c,d,g,h}) pulse sequences are shown. The maps are normalized to the experimental AP$_{2,1}$ \SI{-170}{\per\centi\metre} amplitude and the theoretical AP \SI{-170}{\per\centi\metre} map in panel \textbf{e} is scaled by a factor of 0.5 for clarity.}
	\label{fig:beats}
\end{figure*}

While useful information about the system can be inferred from single-point kinetic traces, coherence signals from multichromophore systems are complex, leaving such an approach insufficient for unambiguous characterization. Fortunately, the complex-valued 2DES datasets contain the signatures of all Liouville (interaction) pathways supported by the laser spectrum\cite{Mukamel_1}. As such -- after subtraction of appropriate multi-exponential population dynamics -- a Fourier transform along the population time, $t_2$, yields $\tilde{\nu}_{2}$ oscillation maps, showing the QB Fourier amplitude dependence on excitation- and detection- energy coordinates \cite{Li_2013,Seibt_2013}. QBs of different origins produce distinct patterns and therefore they can be used to unambiguously determine the origin of coherences \cite{Butkus_2017}.

Such oscillation maps were constructed from both experimental and simulated datasets, and are presented in Fig.~\ref{fig:beats}. In the top row the experimental AP and DC oscillation maps at the dominant $\pm$ \SI{170}{\per\centi\metre} wavenumbers are shown, while the corresponding simulated maps are presented in the bottom row. The similar, but weaker, experimental oscillation maps at $\pm$ \SI{210}{\per\centi\metre} are shown in Supplementary Fig.~4.

The negative wavenumber experimental AP map (Fig.~\ref{fig:beats}a) is richest in structure, particularly in the area around AP$_{2,1}$ where long-lived QBs were reported in earlier studies \cite{Engel_07a, Panitchayangkoon_10a}. Whilst the dephasing times found at AP$_{2,1}$ are indeed in agreement with Ref.~\cite{Panitchayangkoon_10a} (see Fig.~\ref{fig:dyn} and Supplementary Table~1), it is apparent that the observed \SI{-170}{\per\centi\metre} QB is not an isolated feature but rather forms part of a square pattern of below-diagonal peaks. Following the analysis of Butkus \textit{et al.} \cite{Butkus_12a}, this is a characteristic signature of vibrational wavepackets on the electronic ground state.

In stark contrast to this, the AP positive wavenumber map (Fig.~\ref{fig:beats}b) is almost featureless, exhibiting only poorly defined areas of QB amplitude in much of the diagonal and above-diagonal regions. It is thus clear that the AP experimental results are dominated instead by negative wavenumber QBs (see also Fig.~\ref{fig:dyn}b).

The experimental DC oscillation maps (Fig.~\ref{fig:beats}c and d) are not quite different from the AP maps. Two particularly notable differences are the relative sparsity of features, and that the positive- and negative- wavenumber maps are of comparable amplitude. The negative wavenumber DC map features a relatively high-amplitude peak at DC$_{2,1}$, while the corresponding positive wavenumber map has significant amplitude only around DC$_{1,2}$. It has been pointed out elsewhere\cite{Butkus_2017,Thyrhaug_2017} that above-diagonal, positive wavenumber QBs in rephasing spectra, as observed in Fig.~\ref{fig:beats}d, are characteristic of excited state coherence. That this feature does not contribute significantly in the AP experiment (Fig.~\ref{fig:beats}b) may be due to interference with strongly overlapping excited state absorption in the same spectral region.

The simulated negative wavenumber oscillation maps show reasonable agreement with the experiment, where the major features in both AP and DC experiments are well captured, as seen in Fig.~\ref{fig:beats}a,e and c,g. The main discrepancy is the presence of non-negligible features at higher detection energies in the simulated data. Two factors likely contribute to this; (i) the model does not reproduce perfectly the relative exciton oscillator strength, resulting in an apparent signal increase for coherences involving higher energy transitions, and (ii) vibrations on all sites are assumed to have the same Huang-Rhys factor in the model, which results in additional below-diagonal square peak arrangements appearing at higher energies. This assumption is unlikely to be valid, as vibrations involving exciton 2 dominate the experimental coherence response, while vibrations from higher-energy excitons contribute only weakly. We speculated that Herzberg-Teller coupling -- a dependence of the BChl's transition dipoles on nuclear coordinates\cite{Orlandi_73a} -- may play a role due to its influence on excited state displacements. 

Finally, just as with the single-point traces (Fig.~\ref{fig:dyn}b), the excited state coherence expected in the positive wavenumber DC oscillation map appears with too low an amplitude in simulations (Fig.~\ref{fig:beats}h). This suggests that vibronic coupling effects in the excited state may not have been sufficiently accounted for and may be a fruitful area for future improvements in the modelling of the FMO complex.

\section*{Discussion and Conclusions}

Much of the recently published work on FMO has been formulated within a framework of excitonic coherence. This interpretation has however been difficult to reconcile with theoretical models using realistic parameters. Here we show that all long-lived coherence signals in FMO can be explained entirely within a generally well-accepted vibronic coupling framework.

Unambiguous characterization of the convoluted coherence dynamics observed in complex multi-chromophore systems first requires separation of the total measured signal into rephasing and non-rephasing spectra. As demonstrated above, a second effective step is the separation of QBs into positive and negative wavenumbers. This further discrimination is extremely useful as -- in the absence of coherence shifts\cite{ETICS} -- rephasing/non-rephasing ground-state coherences only appear at negative/positive wavenumbers (see Supplementary Fig.~5 for details). Excited state coherences on the other hand, exhibit as a general characteristic equal amplitude contributions of both signs \cite{Butkus_2017,Thyrhaug_2017}. Considerations of the contributing Liouville pathways and the signal sign thus allows identification of the characteristic spectroscopic `fingerprints' of different coherences.

The strongest overall QB contribution (for both 170 and \SI{210}{\per\centi\metre} modes) is the negative-wavenumber, below-diagonal square in the AP experiment. This is characteristic of ground-state vibrations in a displaced oscillator -- \textit{i.e.} it is fully consistent with simple \textit{intramolecular}  vibrations only \cite{Butkus_12a,Butkus_2017}. This interpretation is supported by simulation, where such an arrangement is reproduced as a result of ground state coherence.  Conversely, it is entirely inconsistent with excited state (either excitonic or vibrational) coherence. As a consequence, we can exclude significant contributions from long-lived excitonic coherences in both our experiment and earlier measurements of the FMO complex. Instead, we observe excitonic coherences as early-time QBs dephasing on a $\sim$\SI{100}{\femto\second} time scale (Fig.~\ref{fig:dyn}a). 

The AP experiment is limited in that it does not, \textit{a priori}, reveal much information about the nature of the excited states involved in generating given coherences. In this regard, the DC experiment is more illuminating; the absence of QB signals would designate the involvement of only `trivial' vibrational coherences.  However, as can be seen in Fig.~\ref{fig:dyn} and Fig.~\ref{fig:beats} there are strong QB signals present in DC experiment, which indicates effects of vibronic coupling (mixing) of excited states. 

As QBs of both positive and negative wavenumber contribute in the DC experiment, excited state coherences have to be present\cite{Butkus_2017,Thyrhaug_2017}. This analysis can be extended; positive wavenumber QBs in rephasing spectra require the involvement of excited state processes. They cannot originate from electronic coherences as the dephasing-time of this positive wavenumber contribution ($\tau^{QB} = 570$\si{\femto\second}) is substantially longer than the elecronic coherence lifetime of 50--150\, fs (vide supra). It is however consistent with the predicted signal from the (largely) localized excited state vibrations proposed in the vibronic exciton model by Christensson \textit{et al.}\cite{Christensson_12a}. Thus our measurements provide unambiguous experimental observation of long-lived excited state vibronic coherence in FMO. 

The excited state cannot be the only contributor in the DC experiment however, as the symmetry between negative and positive oscillation maps is limited\cite{Butkus_12a}. While excited state coherence does produce pairs of cross-peak features (such as at DC$_{1,2}$ and DC$_{2,1}$), these -- because of the symmetry of the involved Liouville pathways -- necessarily appear with equal amplitudes and dephasing times. We on the other hand observe a substantially higher QB amplitude and longer average dephasing time at DC$_{2,1}$. This strong, negative wavenumber feature (Fig.~\ref{fig:beats}c) corresponds to one corner of the square in the AP experiment and has a similar dephasing time of $\sim$\SI{2}{\pico\second}. Thus the strongest contribution to the signal around DC$_{2,1}$ appears to originate from the ground-state.
 
Following the analysis of  Jonas and co-workers\cite{Tiwari_13a} we find the clear presence of ground state vibrational coherence, which is enhanced via the vibronically coupled excitonic level at 12,270 cm$^{-1}$ (See Supplementary Fig.~6 for details). The simulated oscillation maps support this assignment, as the experimentally observed negative wavenumber QB at DC$_{2,1}$ -- originating mostly from the ground state -- is successfully reproduced.

While all observed coherences in FMO are thus explained, the absence of higher wavenumber ($>$\SI{210}{\per\centi\metre}) vibrational modes prominent in florescence line narrowing experiments\cite{Ratsep_07a} is notable. It is possible that the dominance of low-frequency modes is due to Herzberg-Teller coupling \cite{Orlandi_73a} which acts to modulate the effective Huang-Rhys factors in coupled systems. Strong enhancement of low wavenumber QBs have been observed in porphyrin aggregates\cite{Kano_2002a} and in chlorosomes \cite{Dostal_14a}. In the latter work, simple modelling showed that the Herzberg-Teller effect preferentially enhances the Huang-Rhys factor of low wavenumber modes.

Work on long-lived QBs in FMO \cite{Engel_07a} has been ongoing for a decade and has given rise to heated debate about the role of quantum coherence in photosynthesis. In this contribution we have demonstrated that polarization-controlled 2DES, aided by theoretical modelling, allows for a clear interpretation of the rich coherence dynamics of the FMO complex. The early times are dominated by short-lived superpositions of excitonic states induced by the broadband laser pulses. Following this, long-lived vibrational coherences of predominantly ground-state origin slowly dephase over approximately \SI{2}{\pico\second}. Through their modulation of the transition polarization, these coherences reveal vibronic mixing in the electronic structure, whose impact on energy transfer is yet to be elucidated. We conclude that the long-lived QBs previously assigned to excitonic coherence are ground-state vibrational in origin -- a finding which is line with recent theoretical work \cite{Tiwari_13a, Tempelaar_14a}. While excitonic coherences are observed, they decay completely in less than \SI{240}{\femto\second} at the cryogenic temperatures used here. This coherence was also notably absent in room-temperature 2DES experiments \cite{Duan_2017}, suggesting even faster dephasing under physiological conditions. Such a short life-time strongly suggests that an electronic coherence does not contribute to energy transfer in FMO, even under the speculative assumption that such a coherence can indeed be prepared by energy transfer from the chlorosome under natural conditions.
 
We believe the demonstrated approach will prove useful for disentangling the complex coherence signals of other multi-chromophore light-harvesting complexes, and thus provide a stimulus to ongoing research aimed at unravelling the role of vibronic coupling in energy transfer within photosynthetic systems.

\section*{Methods}

\textit{C. tepidum culture and FMO preparation}. 
\textit{C. tepidum} (strain TLS, DSM 12025) was grown in 800 mL batches of modified Pfennig medium \cite{Wahlund_1991} in 1 litre flasks immersed in temperature-controlled water baths at 45°C. Continuous illumination was provided by 60 W incandescent light bulbs. Protein purification was based on Wen et al. \cite{Wen_2009}. The isolation buffer for all purification steps was 20 mM Tris-HCl, pH 8. The cells were harvested after 3 days of cultivation, by centrifugation at 6000 g, resuspended and broken using EmulsiFlex C5 (Avestin Inc., Canada) at 20000 psi. Unbroken cells were removed by low speed centrifugation and the membrane fragments present in the resulting supernatant were collected by ultracentrifugation at 200000 g for 2 hours and then resuspended in isolation buffer. FMO was released from membranes by 0.4 M Na2CO3 added in two steps over the course of 2 days (at 4$^\circ$ C in the dark) to release FMO. The soluble protein fraction was cleared of debris by ultracentrifugation, dialysed against the isolation buffer for 72 hours, concentrated and purified using size exclusion and anion exchange chromatography until OD271 / OD371 ratio decreased below 0.6. Prior 2DES experiments the sample was dissolved in a 2:1 glycerol:buffer solution, and was held at \SI{77}{\kelvin} in a nitrogen flow cryostat during the entire experiment.

\textit{Experiment}. 
Details of our implementation of non-collinear 2DES with the double-frequency lock-in detection scheme can be found in \cite{Augulis_11a}. Briefly, three ultrashort laser pulses interact with a sample, inducing a third-order polarization. The radiation of the polarization is heterodyne detected using a fourth pulse (the local oscillator) by measuring spectrally dispersed interferograms with a CCD camera. The detected signal is Fourier transformed with respect to the time interval between the first and second pulses, yielding the conjugate excitation frequency, and with respect to the interval between the third pulse and signal, yielding the detection frequency. The remaining independent time interval is the pump-probe delay analogue and is referred to as the population time. Fourier transformed data are represented as two-dimensional excitation-detection spectra obtained at a given population time. Broadband femtosecond pulses -- approximately \SI{100}{\nano\metre} in bandwidth and centred at \SI{805}{\nano\metre} -- were generated by a home-built non-collinear optical parametric amplifier seeded by the 1030 nm output of a Yb:KGW amplified laser system (Pharos, Light Conversion). The resulting pulses were compressed to \SI{14}{\femto\second} using a combination of chirped mirrors and a prism-pair compressor. Two pairs of phase-stable pulses were generated by first splitting the NOPA output pulses with a beamsplitter and subsequently focusing them on a diffraction grating optimized for the $\pm 1$ orders, further splitting the two pulses into two phase-locked pairs. The linear polarization of the all four pulses was independently controlled by a combination of the quarter-wave plate and linear wire-grid polarizers placed in each beam.  The population time-delay between the pulse pairs was controlled by an optical delay line, while the coherence time-delay between pulses 1 and 2 were controlled by pairs of fused silica wedges. The coherence time was scanned in 1.8 fs steps until the signal decayed into noise. This corresponded to a range of -270 to 450\,fs in AP experiments and -130 to 230\,fs  in DC experiments. The resulting spectral resolutions on the excitation axis after Fourier transformation over the coherence time were 36 and \SI{72}{\per\centi\metre} respectively, whereas resolution on the detection axis was \SI{40}{\per\centi\metre} in both experiments. The population time was scanned in regular 20\,fs steps from 0 to 2 ps, and the signal-to-noise ratio was increased by averaging two consecutive scans for the AP dataset and five scans for the DC dataset. No zero-padding of population time data was used for quantum beat analysis. The estimated suppression of the signals involving interactions with pairwise parallel dipoles in the DC measurement is $\sim$ 125 times. 

\textit{Modelling}. Calculations were based on earlier electronic parametrizations of the Fenna-Matthews-Olson (FMO) complex from \textit{Chlorobium tepidum}\cite{Vulto_98a, Brixner_05a, Tronrud2009}. The quantum-mechanical degrees of freedom were described using the Holstein Hamiltonian\cite{Holstein_59a}, which explicitly accounts for linear coupling of the electronic transition to a single vibrational mode for each BChl. In doing so, a vibrational wavenumber of \SI{160}{\per\centi\metre} was used in the calculations. The remaining modes were described using the over-damped Brownian oscillator model\cite{Mukamel_1}, using a fluctuation width and time scale consistent with molecular dynamics simulations\cite{Olbrich_11a}. The quantum dynamics was simulated through numerical integration of the Schr\"odinger equation\cite{Jansen_06a, Torii_06a}, neglecting feedback of the quantum degrees of freedom onto the (classical) modes in the environment (effectively adapting an asymptotic high-temperature approximation for the quantum system). Absorption and 2D spectra were obtained through evaluation of the 2-point and 4-point dipole correlation functions, respectively\cite{Mukamel_1}. The absorption spectrum was averaged over 50,000 (uncorrelated) bath trajectories, whereas for the 2D spectra an average over 250,000 trajectories were taken (for details see Supplementary Section 1).

\section*{Acknowledgements}
The work in Lund was supported by the Swedish Research Council, the Knut and Alice Wallenberg Foundation and the Crafoord Foundation. R.T. acknowledges The Netherlands Organisation for Scientific Research NWO for support through a Rubicon grant. D. B. acknowledges funding from Czech Science Foundation under the grant number P501/12/G055 and institutional support RVO:60077344. We thank Dr. David Pale\v{c}ek for making his code for complex QB analysis available.

\section*{Author Contributions}
D.Z. conceived the idea, E.T., M.A., K.Z. and D.Z. designed and performed experiments, R.T., J.K. and T.L.C.J designed the theory, R.T. performed simulations, and D.B. extracted and purified the sample. E.T. M.A. and R.T. analysed the data. E.T., R.T. and D.Z. wrote the manuscript with input from all the other authors.

\section*{Competing financial interests}
The authors declare no competing financial interests.



\end{document}